\documentclass[aps,prl,twocolumn,showpacs]{revtex4}
\usepackage[dvips]{graphicx}
\usepackage{epsf}
\usepackage{dcolumn}
\usepackage{bm}
\usepackage{amsmath}
\usepackage{amssymb}
\usepackage[dvipsnames]{xcolor}
\usepackage{color} 
\usepackage{ulem}

\usepackage[dvipdfmx,
colorlinks=true,
linkcolor=magenta,
urlcolor=magenta,
citecolor=magenta,
filecolor=blue,
pagecolor=blue,
linktocpage=true,
bookmarks=true,
bookmarksnumbered=true
]{hyperref}

\begin{document}


\title{ 
Spin-Orbital Entangled Excitonic Insulators in $(t_{2g})^4$ Correlated Electron Systems
}

\author{Toshihiro Sato$^1$}
\author{Tomonori Shirakawa$^{2}$}
\author{Seiji Yunoki$^{1,2,3}$}

\affiliation{
$^1$Computational Condensed Matter Physics Laboratory, RIKEN, Wako, Saitama 351-0198, Japan\\
$^2$Computational Quantum Matter Research Team, RIKEN Center for Emergent Matter Science (CEMS), Wako, Saitama 351-0198, Japan\\
$^3$Computational Materials Science Research Team, RIKEN Advanced Institute for Computational Science (AICS), Kobe, Hyogo 650-0047, Japan
}

\date{\today}

\begin{abstract}

We employ the multi-orbital dynamical mean-field theory to examine the ground state of a 
three-orbital Hubbard model with a relativistic spin-orbit coupling (SOC) at four electrons per site. 
We demonstrate that the interplay between the strong electron correlations and the SOC induces 
a Van Vleck-type nonmagnetic insulator and its magnetic exciton condensation. 
We also find in the moderate electron correlation regime that the SOC induces another type of 
a nonmagnetic excitonic insulator, in addition to a relativistic band insulator. 
The characteristic features among these insulators are manifested in the 
momentum resolved single-particle excitations, thus accessible by 
angle-resolved photoemission spectroscopy experiments.

\end{abstract}


\maketitle

Since the observation of an antiferromagnetic (AF) insulator in $5d$ transition metal Ir oxides 
$A_{2}{\rm IrO}_{4}$ ($A$ = Sr and Ba)~\cite{Randall,Cava,Shimura,Cao,Okabe}, 
many properties of the insulator have been explored both theoretically and experimentally.
These materials are crystalized in a layered perovskite structure with nominally five $5d$ electrons 
per Ir in $t_{2g}$ orbitals which are separated from $e_g$ orbitals due to a large crystal field.
The important difference from $3d$ and $4d$ transition metal oxides
is that $5d$ transition metal oxides show a strong relativistic spin-orbit coupling (SOC) along with 
the moderate electron correlations.
Because of the strong SOC, $t_{2g}$ orbitals are split into the local effective total angular momentum 
$j=1/2$ doublet and $j=3/2$ quartet in the atomic limit~\cite{Sugano}.
The theoretical and experimental studies have revealed 
the AF insulating ground state with the half-filled $j=1/2$ based band, where the $j=3/2$ based bands are completely occupied
~\cite{Kim1,Kim2,Ishii,Fujiyama,Jackeli,Jin,Watanabe1,Shirakawa,Martins,Arita,Onishi,Watanabe3}.
The low temperature phase diagram in the $(t_{2g})^5$ electron system has also been reported theoretically, focusing on the competition between the electron correlations and the SOC~\cite{Sato1}.
It has been also predicted that the carrier doping can induce unconventional 
superconductivities~\cite{Wang2,Watanabe2,Yang,Meng}

Recently, two experiments have reported interesting observations for perovskite and 
post-perovskite $5d$ transition metal oxides, 
$A{\rm OsO}_{3}$ ($A$ = Ca, Sr, and Ba)~\cite{Shi,Zheng} 
and $\rm NaIrO_{3}$~\cite{Bremholm}, respectively, 
with a nominally $(t_{2g})^4$ electron configuration. 
%
At low temperatures, all these materials except for $\rm SrOsO_{3}$ exhibit an insulating behavior 
with no indication of magnetic order, 
despite that the first-principles calculations 
based on density functional theory (DFT), including LDA(GGA)+SOC+$U$ calculations, 
predict that these materials are all metallic~\cite{Shi,Bremholm,Du2,Jung}. 
%
%
%

When the SOC is significantly large, i.e., in the $jj$ coupling limit, the $j=1/2$ and $3/2$ based 
bands are well separated 
to simply become a relativistic band insulator with the fully occupied 
$j=3/2$ based bands and the completely empty $j=1/2$ based band. 
%
Indeed, Du $et~al.$ has performed the LDA+Gutzwiller calculation, which combines the DFT with the local density 
approximation (LDA) and the Gutzwiller variational method, and concluded that $\rm NaIrO_{3}$ is such a 
correlated paramagnetic band insulator~\cite{Du2,Du1}. 

%
On the other hand, when the electron correlations are dominant, 
the $LS$ coupling scheme in the atomic limit is a better description.  
%
According to the Hund's rule, the ground state of a $(t_{2g})^4$ electron system 
has the orbital and spin angular momenta $L=1$ and $S=1$, respectively, which couple via the SOC to form 
the total angular momentum $J=0$. 
%
Considering the hybridization between neighbors, it is expected that the local singlets with $J=0$ form 
a band to become 
a Van Vleck-type nonmagnetic insulator.
%
In this context, Khaliullin has recently proposed a Van Vleck-type excitonic insulator, 
where an excitonic condensation between the local $J=0$ and $J=1$ states is driven by the 
magnetic order through the intersite exchange interaction~\cite{Khaliullin}.
%



The theoretical studies in the two extreme limits are important to investigate possible exotic states.
However, it is highly desirable to examine the electronic ground state as well as the 
excitations in a wider range of couplings using a numerical technique which allows us to treat 
the interplay between the electron correlations and the SOC in a well controlled 
manner. 

For this purpose, here
we employ the multi-orbital dynamical mean field theory (DMFT)~\cite{MODMFT} to 
study a three-orbital Hubbard model with the SOC at four electrons per site,
corresponding to the $(t_{2g})^4$ electronic configuration. 
%
%
We demonstrate that the interplay between the strong electron correlations and the SOC can stabilize 
the Van Vleck-type nonmagnetic insulator 
and its magnetic exciton condensation. 
Moreover, we find in the moderate electron correlation regime that the SOC can induce a nonmagnetic 
excitonic insulator formed by an electron-hole pair in the $j=1/2$ and $3/2$ based bands, 
as well as the relativistic band insulator. 
The characteristic features among these different insulators are found in the momentum 
resolved single-particle excitations.

The three-orbital Hubbard model with the relativistic SOC studied here is given as 
$H=H_0+H_{\rm I}$, where 
%
$
H_0=\sum_{\langle i,i^{\prime} \rangle}\sum_{\gamma,\sigma}t^{\gamma}c_{i\gamma\sigma}^{\dagger} c_{i^{\prime}\gamma\sigma}
-\mu \sum_{i,\gamma,\sigma} n_{i\sigma}^\gamma
+\lambda \sum_i\sum_{\gamma,\delta}\sum_{\sigma,\sigma'}\langle \gamma|\mathbf l_{i}|\delta \rangle \cdot \langle \sigma|\mathbf s_{i}|\sigma' \rangle c_{i\gamma\sigma}^{\dagger} c_{i\delta\sigma'}
$
and
$
H_{\rm I}=U\sum_{i,\gamma}n_{i\uparrow}^\gamma n_{i\downarrow}^\gamma+\frac{U'-J_z}{2}\sum_{i,\gamma\neq\delta,\sigma}n_{i\sigma}^\gamma n_{i\sigma}^\delta 
+\frac{U'}{2}\sum_{i,\gamma\neq\delta,\sigma}n_{i\sigma}^\gamma n_{i\bar{\sigma}}^\delta
-J_{xy}\sum_{i,\gamma\neq\delta}c_{i\gamma\uparrow}^{\dagger} c_{i\gamma\downarrow} c_{i\delta\downarrow}^{\dagger}  c_{i\delta\uparrow} 
+J_{xy}\sum_{i,\gamma\neq\delta}c_{i\gamma\uparrow}^{\dagger}  c_{i\gamma\downarrow}^{\dagger}  c_{i\delta\downarrow} c_{i\delta\uparrow}
$. 
$H_0$ represents the non-interacting part, where $t^{\gamma}$ sets the nearest-neighbor hopping amplitude 
for $t_{2g}$ orbitals $\gamma=(d_{yz}, d_{zx}, d_{xy})$ and 
$\mu$ is the chemical potential tuned to be at four electrons per site.  
$\lambda$ is the SOC and $\mathbf l_{i}$ ($\mathbf s_{i}$) is the orbital (spin) angular momentum operator 
at site $i$. 
$ c_{i\gamma\sigma}^{\dagger}$ is an electron creation operator at site 
$i$ with orbital $\gamma$ and spin $\sigma\,(=\uparrow,\downarrow)$
and $n_{i\sigma}^{\gamma}=c_{i\gamma\sigma}^{\dagger} c_{i\gamma\sigma}$. 
The local interacting part $H_{\rm I}$ includes the intra (inter) orbital Coulomb interaction $U$ ($U'$), 
the Hund's rule coupling $J_z$, and the spin flip and pair hopping $J_{xy}$,
where we set $U=U'+2J_z$ and $J_z=J_{xy}=0.15U$~\cite{Kanamori}.
The electron correlations as well as ordered states are treated by the multi-orbital DMFT on the Bethe lattice 
with coordination number $Z \to \infty$~\cite{Eckstein,Metzner}, where the DMFT is exact.
We consider the same bandwidth $W$ for the three orbitals ($t^{\gamma}=t/\sqrt{Z}$), i.e., 
$W=4t$, and $t$ is used as the energy unit. 
In the following, we show the results for the lowest temperature $T=0.06$, which essentially represent 
the ground state.

\begin{figure}
\centering
\vspace{0cm}
\centerline{\includegraphics[width=1\hsize]{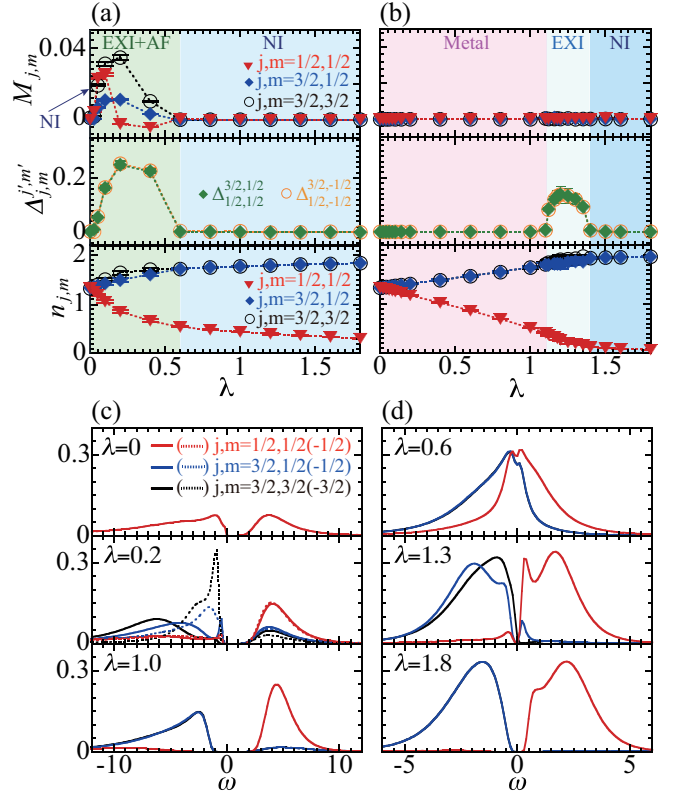}
}
\caption{(Color online) (a) and (b): $\lambda$ dependence of staggered magnetization $M_{j,m}$ (top), 
excitonic order parameter $\Delta_{j,m}^{j',m'}$ (middle), and electron density $n_{j,m}$ (bottom). 
(c) and (d): Single-particle excitations spectrum $A_{j,m}(\omega)$ for three different $\lambda$'s.  
Note that, except for $\lambda=0.2$ in (c), all components of the spectra are degenerate for 
$\lambda=0$ in (c), $A_{j,m}(\omega)=A_{j,-m}(\omega)$ for $\lambda=1.3$ in (d), and 
$A_{3/2,\pm3/2}(\omega)=A_{3/2,\pm1/2}(\omega)$ and $A_{j,m}(\omega)=A_{j,-m}(\omega)$ for other three 
parameters. 
We set $U=12$ [(a) and (c)] and $3$ [(b) and (d)]. 
EXI, EXI+AF, and NI stand for excitonic insulator, antiferromagnetic excitonic insulator, and nonmagnetic 
insulator, respectively. 
}  
\label{fig:op}
\end{figure}

In the multi-orbital DMFT calculation, we numerically obtain the
imaginary-time Green's functions in the impurity site
$G_{\gamma,\sigma}^{\delta,\sigma'}(\tau) \equiv -\langle T_{\tau}c_{i\gamma \sigma}(\tau)c_{i\delta \sigma'}^\dagger(0)\rangle $
by using a continuous-time quantum Monte Carlo (CTQMC) method based on the strong coupling 
expansion~\cite{CTQMC}.
Although the CTQMC calculation in principle enables us to solve the model exactly, the negative sign problem 
is one of the serious issues, particularly at low temperatures for large SOC.
Our previous study for $(t_{2g})^5$ correlated systems have demonstrated that the sign problem is improved 
by transforming the $t_{2g}$ orbital bases ($c_{i\gamma\sigma}$) to the maximally spin-orbit-entangled
$j$ bases ($a_{ijm}$) of the eigenstates of $H_0$ in the atomic limit, i.e., 
%
%
\begin{eqnarray}
\left(
    \begin{array}{c}
      a_{i\frac{1}{2}\frac{s}{2}} \\
      a_{i\frac{3}{2}\frac{s}{2}} \\
      a_{i\frac{3}{2}\frac{-3s}{2}}
    \end{array}
  \right)
=\frac{1}{\sqrt{6}}
\left(\begin{array}{ccc}
\sqrt{2} & -i\sqrt{2}s & \sqrt{2}s \\
s & -i & -2 \\
 -\sqrt{3}s & -i\sqrt{3} & 0 \\
\end{array} 
\right)
\left(
    \begin{array}{c}
      c_{id_{yz}\bar\sigma} \\
      c_{id_{zx}\bar\sigma} \\
      c_{id_{xy}\sigma}
    \end{array}
  \right), \nonumber
\end{eqnarray}
where $s=1\,(-1)$ for $\sigma=\uparrow(\downarrow)$ and $\bar\sigma$ implies the opposite spin 
to $\sigma$~\cite{Sato1,note1}. 
We find that the same basis transformation can improve the sign problem also for the present 
$(t_{2g})^4$ systems. 
%

We investigate two ordered states attributed to the interplay between the electron correlations and 
the SOC. 
The first state considered is the magnetic order along the $z$ direction described by the order parameter  
$M_{j,m}(l)=\frac{1}{2}\sum_{m'=\pm m}{\rm sign}(m')\langle a_{ljm'}^{\dagger} a_{ljm'}\rangle$, 
%
where $l(= A,\,B)$ indicates two sublattices~\cite{note4}.
We also introduce the excitonic order parameter formed in 
different $j$ orbitals, i.e., 
$\Delta_{j,m}^{j',m'}(l)=\langle a_{ljm}^{\dagger} a_{lj'm'}\rangle$, 
where $j\ne j'$. 
%
In addition, we calculate the electron density
$n_{j,m}(l)=\sum_{m'=\pm m}\langle a_{ljm'}^{\dagger} a_{ljm'}\rangle$. 
Since we have found that $M_{j,m}(A)=-M_{j,m}(B)$, implying AF order, 
$\Delta_{j,m}^{j',m'}(A)=-\Delta_{j,-m}^{j',-m'}(B)$, and $n_{j,m}(A)=n_{j,-m}(B)$, 
we omit the sublattice index $l$ in these quantities below.

\begin{figure}
\centering
\vspace{0cm}
\centerline{\includegraphics[width=1\hsize]{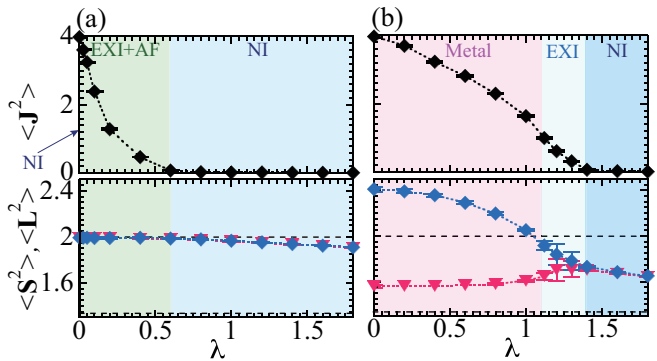}
}
\caption{(Color online)
$\lambda$ dependence of total angular momentum sqaured $\langle \mathbf J^2 \rangle$ (top) and 
local spin (orbital) angular momentum squared $\langle  \mathbf S^2 \rangle$ 
($\langle \mathbf L^2 \rangle$) (bottom) for (a) $U=12$ and (b) $U=3$.
Dashed lines in bottom panels indicate $\langle  \mathbf S^2 \rangle=\langle \mathbf L^2 \rangle=2$ and 
$4/3$. 
EXI, EXI+AF, and NI stand for excitonic insulator, antiferromagnetic excitonic insulator, and nonmagnetic 
insulator, respectively. 
}  
\label{fig:JSL}
\end{figure}

We first examine a case with large $U$. 
Figure~\ref{fig:op}(a) shows $M_{j,m}$, $\Delta_{j,m}^{j',m'}$, and $n_{j,m}$ for $U=12$ as a function of $\lambda$. 
When $\lambda=0$, there are no magnetic and excitonic orders with 
$n_{j,m}=4/3$.
However, as soon as $\lambda$ is finite, 
the excitonic order parameters become finite with $\Delta_{1/2,1/2}^{3/2,1/2}=\Delta_{1/2,-1/2}^{3/2,-1/2}\ne0$.
%
Concomitantly, the magnetic order parameters are also finite for all three $j$ components and 
are antiferromagnetically aligned. 
%
%
%
This phase extends up to $\lambda=0.6$. 
%
It should also be noticed that $n_{3/2,1/2}\ne n_{3/2,3/2}$ in the presence of a finite excitonic order~\cite{note3}.  
%
For $\lambda\ge0.6$, both magnetic and excitonic orders disappear, 
%
and $n_{3/2,1/2}=n_{3/2,3/2}$ ($n_{1/2,1/2}$) increases (decreases) towards two (zero) with further 
increasing $\lambda$. 
%
In Fig.~\ref{fig:op}(c), we calculate the single-particle excitation spectrum
$A_{j,m}(\omega)=-\frac{1}{\pi} {\rm Im}G_{j,m}^{j,m}(\omega+i0)$, 
where $G_{j,m}^{j',m'}(\tau) \equiv -\langle T_{\tau}a_{ijm}(\tau)a_{ij' m'}^\dagger(0)\rangle $ and 
$\omega$ is real frequency, 
and find that there is a finite gap at the Fermi 
energy for all values of $\lambda$. Therefore, these states are all insulating.

To understand the nature of these insulating states, we examine the local spin and orbital angular 
momenta squared, $\langle \mathbf S_i^2 \rangle$ and $\langle \mathbf L_i^2 \rangle$, respectively, 
and 
the local total angular momentum squared $\langle \mathbf J_i^2 \rangle$, 
where 
$\mathbf S_i = \sum_{\gamma}\sum_{\sigma,\sigma'} \langle \sigma|\mathbf s_{i}|\sigma' \rangle c_{i\gamma\sigma}^{\dagger} c_{i\gamma\sigma'}$, 
$\mathbf L_i = \sum_{\gamma,\delta}\sum_{\sigma}\langle \gamma|\mathbf l_{i}|\delta \rangle  c_{i\gamma\sigma}^{\dagger} c_{i\delta\sigma}$, and 
$\mathbf J_i=\mathbf S_i-\mathbf L_i$. 
Since these quantities do not depend on the site index $i$, we simply omit this index hereafter. 
First, it is highly instructive to consider four limiting cases. According to the Hund's rule, 
when the electron correlation is significantly large with no SOC, we expect that 
$\langle \mathbf S^2 \rangle=\langle \mathbf L^2 \rangle=2\ {\rm and}\ \langle \mathbf J^2 \rangle=4$
because $\langle \mathbf L\cdot \mathbf S\rangle=0$. If the SOC is introduced in this limit, the spin and 
orbital angular momenta are aligned antiparallel, and thus 
$\langle \mathbf S^2 \rangle=\langle \mathbf L^2 \rangle=2\ {\rm and}\ \langle \mathbf J^2 \rangle=0. $
This is exactly the case for the ideal Van Vleck-type insulator. 
In the limit of significantly large SOC, where the ground state is expected to be 
the relativistic band insulator, 
$\langle \mathbf S^2 \rangle=\langle \mathbf L^2 \rangle=\frac{4}{3}\ {\rm and}\ \langle \mathbf J^2 \rangle=0.$
Finally, in the noninteracting limit without the SOC, 
$\langle \mathbf S^2 \rangle=1.2,\ \langle \mathbf L^2 \rangle=3.2,\ {\rm and}\ \langle \mathbf J^2 \rangle=4.4.$

%
Figure~\ref{fig:JSL}(a) shows the evolution of these quantities 
with varying $\lambda$ for $U=12$. 
As expected in the strong electron correlation limit, we find that 
$\langle \mathbf S^2 \rangle=\langle \mathbf L^2 \rangle=2$ and $\langle \mathbf J^2 \rangle=4$ 
for $\lambda=0$, in accordance with the Hund's rule. 
With increasing $\lambda$, $\langle \mathbf J^2 \rangle$ monotonically decreases 
and eventually becomes zero for $\lambda\ge0.6$, where neither magnetic nor excitonic order exists, 
as shown in Fig.~\ref{fig:op}(a). In this region, $\langle \mathbf S^2 \rangle$ and $\langle \mathbf L^2 \rangle$ 
are still close to two, specially near $\lambda=0.6$, implying that the nonmagnetic insulator for $\lambda\ge0.6$ 
is the Van Vleck-type insulator. However, we also notice that $\langle \mathbf S^2 \rangle$ and 
$\langle \mathbf L^2 \rangle$ gradually decreases with further increasing $\lambda$, while 
$\langle \mathbf J^2 \rangle$ is exactly zero, and we expect that $\langle \mathbf S^2 \rangle$ and 
$\langle \mathbf L^2 \rangle$ eventually become $4/3$ in the limit of $\lambda\to\infty$. 
Namely, the Van Vleck-type insulator 
is smoothly connected to the simple relativistic band insulator. 
On the other hand, we find in the region of $0<\lambda<0.6$ that $\langle \mathbf J^2 \rangle>0$ 
while $\langle \mathbf S^2 \rangle$ and $\langle \mathbf L^2 \rangle$ remain two. 
In this region, the magnetic and excitonic orders are both finite, and therefore we attribute this phase to 
the magnetic excitonic insulator with the enhanced hybridization between the nonmagnetic $J=0$ state and 
the magnetic $J\ne0$ 
excited states. 


Next, we examine a case with moderate $U=3$.
It is noticed first in Fig.~\ref{fig:JSL}(b) that $\langle \mathbf J^2 \rangle$, 
$\langle  \mathbf S^2 \rangle$, and $\langle \mathbf L^2 \rangle$ for $\lambda=0$ are 
clearly depart from the noninteracting values, implying that the electron correlations are indeed considered 
to be moderated, as it is expected for $U\approx W$. 
As shown in Fig.~\ref{fig:op}(b), 
we find no magnetic order for all values of $\lambda$. 
%
$A_{j,m}(\omega)$ exhibits sharp quasiparticle peaks around the Fermi energy for both 
$j=1/2$ and $3/2$ 
when $\lambda<1.2$ [see Fig.~\ref{fig:op}(d)], 
and thus the ground state for $\lambda<1.2$ is metallic. 
However, the quasiparticle peaks disappear with further increasing $\lambda$ and 
the metal-insulator transition occurs at $\lambda\sim1.2$, where the single-particle excitation 
gap is open. 
%
More interestingly, in the intermediate SOC region for $1.2\le\lambda\le1.4$, we find 
that
the excitonic insulator emerges without magnetic order [see Fig.~\ref{fig:op}(b)]. 
As shown in Fig.~\ref{fig:op}(d), the degeneracy of $A_{j=3/2,m}(\omega)$ for $m=\pm1/2$ and 
$\pm3/2$ is lifted in this phase. 
%
With further increasing $\lambda$, the excitonic order disappears for $\lambda > 1.4$ 
and the $j=3/2$ based bands are almost completely occupied with a finite excitation 
gap to the unoccupied $j=1/2$ based band [see Figs.~\ref{fig:op}(b) and \ref{fig:op}(d)]. 
This implies that the ground state for $\lambda > 1.4$ is the relativistic band insulator. 
%
%
In the noninteracting limit, the SOC drives a transition from the metal to the relativistic band insulator 
at $\lambda=4/3$. Therefore, the excitonic insulator found here is induced by the electron correlations 
just before the overlap between the $j=1/2$ 
and $3/2$ based bands is diminished, where the condensation of an electron-hole pair in the $j=1/2$ and 
$3/2$ based bands is most favorable with nonzero $\Delta_{1/2,\pm1/2}^{3/2,\pm1/2}$. 

%
%

%
Let us now discuss the difference between the two excitonic insulators, 
i.e., the nonmagnetic excitonic insulator for moderate $U$ and the magnetic excitonic insulator for large $U$. 
The former appears between 
the metal and the relativistic band insulator,
where an electron-hole pair in the $j=1/2$ and $3/2$ based bands are condensed. 
Therefore, the exciton condensation is 
similar to the conventional one 
where a valence hole and a conduction electron form a pair. 
In contrast, the other excitonic insulator 
appears in the strong coupling regime where $U$ alone can 
open the single-particle excitation gap without magnetic order (see Fig.~\ref{fig:op} for $\lambda=0$). 
As shown in Fig.~\ref{fig:JSL}(a), in this strong coupling regime, 
the $LS$ coupling scheme is a better description and thus the 
excitonic insulator here, accompanying the magnetic order, 
is considered as the Van Vleck-type excitonic insulator induced by mixing 
the nonmagnetic $J=0$ state and the magnetic $J=1$ and $J=2$ excited states.
This is simply because $\langle  \mathbf J^2 \rangle>2$ is possible only when 
the $J=2$ state is involved. 
Indeed, we can readily show that the finite order parameter 
$\Delta_{1/2,\pm1/2}^{3/2,\pm1/2}$ generates the mixing 
between the $J=0$ singlet and the $J=2$ quintets in addition to the $J=1$ triplets. 
%
%
Therefore, this magnetic excitonic insulator is similar to the one proposed by Khaliullin~\cite{Khaliullin} 
except for the involvement of the $J=2$ quintets.  

\begin{figure}
\centering
\vspace{0cm}
\centerline{\includegraphics[width=1\hsize]{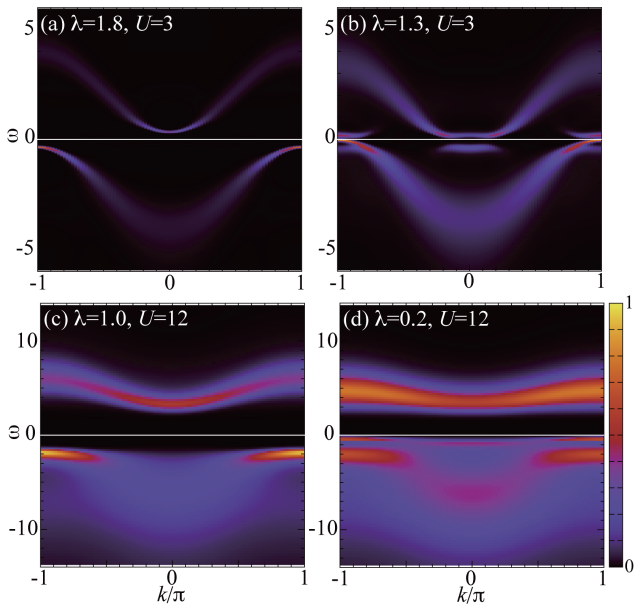}
\vspace{-0.15cm}
}
\caption{(Color online) 
$k$-resolved single-particle spectral function $A(k, \omega)$ for relativistic band insulator, 
(b) nonmagnetic excitonic insulator, (c) Van Vleck-type nonmagnetic insulator, 
and (d) magnetic excitonic insulator.
The maximum of each spectral function is normalized to be one.
The Fermi energy is denoted by horizontal lines at $\omega=0$.
}  
\vspace{-0.5cm}
\label{fig:Ak}
\end{figure}

Finally, we calculate the momentum resolved single-particle spectral function 
%
%
%
$
A(k, \omega)=-\frac{1}{\pi} \sum_{jm}{\rm Im}{\cal G}_{j,m}^{j,m}(k,\omega+i0)
$
and examine the characteristic features in the single-particle excitations for the different insulating 
states found here. 
In our calculations, the single-particle Green's function ${\cal G}_{j,m}^{j,m}(k,\omega)$ 
is introduced within the cluster perturbation theory~\cite{CPT}.
Since the DMFT includes the momentum $k$ dependence 
only through the noninteracting energy dispersion $\epsilon(k)$, we introduce $k$ 
to parametrize $\epsilon(k)$ as 
$\epsilon(k)=-2\cos k$~\cite{Hoshino}. 
%

The typical results of $A(k, \omega)$ are summarized in Fig.~\ref{fig:Ak} for the four different insulators. 
%
%
As shown in Fig.~\ref{fig:Ak}(a), $A(k, \omega)$ for the relativistic band insulator 
in the moderate correlation regime 
is very similar to the noninteracting band structure for large $\lambda$ and exhibits an indirect gap between 
the $j=1/2$ and $3/2$ based bands. 
%
With slightly decreasing the SOC, the ground state becomes the nonmagnetic excitonic insulator and the 
typical result of $A(k, \omega)$ 
is shown in Fig.~\ref{fig:Ak}(b). This result clearly demonstrates 
the strong hybridization between the bottom of the $j=1/2$ based conduction band and the top of 
the $j=3/2$ based valence bands to induce a finite gap at the Fermi energy, and this is a strong evidence for 
the excitonic insulator [see also Fig.~\ref{fig:op}(d)]. 
It is also noticed in Fig.~\ref{fig:Ak}(b) that the spectrum is rather broad, as compared with the one for the 
relativistic band insulator shown in Fig.~\ref{fig:Ak}(a), despite that $U$ is the same for both cases. 

Figure~\ref{fig:Ak}~(c) represents the result for the Van Vleck-type insulator 
which appears in the strong correlation regime. 
The apparent difference from the moderate correlation regime is that the 
single-particle excitation gap is determined by $U$ and the spectrum is rather featureless with a much broader 
structure specially in the unoccupied states. 
%
%
%
The typical result for the magnetic excitonic insulator 
is shown in Fig.~\ref{fig:Ak}(d).
Similar to the Van Vleck-type insulator, 
$A(k, \omega)$ exhibits a broad structure. 
In addition, there exists a characteristic peak structure near $\omega=0$ below the Fermi energy. 
The dispersion of this excitation is strongly renormalized to become almost flat, implying the strong 
correlation effects. 
As shown in Fig.~\ref{fig:op}(c), 
this excitation originates mainly from 
the one with the $m=\pm1/2$ character of $j=3/2$, the fourfold degeneracy of $j=3/2$ being lifted 
due to the magnetic excitonic order.
%

In summary, we have studied the three-orbital Hubbard model with the SOC 
at four electrons per site 
by using the multi-orbital DMFT and the CTQMC method.
For large $U$, we have demonstrated that the moderate SOC induces the Van Vleck-type 
nonmagnetic insulator, which is smoothly connected to the relativistic band insulator for larger $\lambda$. 
We have also found in the strong electron correlation regime that the magnetic excitonic insulator 
%
is induced for small $\lambda$ by hybridizing the nonmagnetic $J=0$ 
singlet of the local $(t_{2g})^4$ manifold and the excited multiplets such as $J=1$ triplets 
and $J=2$ quintets.
%
%
More interestingly, we have found in the moderate electron correlation regime that the excitonic insulator 
emerges without any magnetic order. 
%
This excitonic insulator is due to the condensation of an electron-hole pair in the $j=1/2$ and $3/2$ based 
bands. 
Although our results for moderate $U$ are most appropriate for $5d$ transition metal oxides, 
the results for large $U$ should also be relevant to $3d$ and $4d$ transition metal oxides 
with the low spin configuration. 
The different insulators found here are manifested most clearly in the momentum resolved single-particle 
excitations, and thus can be observed in angle-resolved photoemission spectroscopy experiments. 
 


The authors are grateful to K. Seki for valuable discussion. 
The numerical computations have been performed with facilities at Supercomputer Center in ISSP, 
Information Technology Center, University of Tokyo, 
and with the RIKEN supercomputer system (HOKUSAI GreatWave). 
This work has been supported 
by Grant-in-Aid for Scientific Research from MEXT Japan 
under the Grant No. 25287096 and also in part 
by RIKEN iTHES Project and Molecular Systems.



\begin{thebibliography}{}
\bibitem{note1}
The imaginary unit $i$ in the transformation matrix should not be confused with the site index $i$ 
in the annihilation operators.
\bibitem{note4}
Since our model is isotropic, the magnetic orders along the other $x$ and $y$ directions are identical to that 
along the $z$ direction.
\bibitem{note3}
This is also the case in another excitonic insulating phase for $U=3$.
\end{thebibliography}
\end{document}